\documentstyle[12pt]{article}
\def\lsim{\mathrel{\rlap{\lower4pt\hbox{\hskip1pt$\sim$}}
    \raise1pt\hbox{$<$}}}         %less than or approx. symbol
\def\gsim{\mathrel{\rlap{\lower4pt\hbox{\hskip1pt$\sim$}}
    \raise1pt\hbox{$>$}}}         %greater than or approx. symbol

\def\beq{\begin{equation}}
\def\endeq{\end{equation}}
\def\arr{\begin{eqnarray}}
\def\endarr{\end{eqnarray}}

\makeindex
\begin{document}
%\large
\phantom{.}\hspace{8.0cm} 9 March 1993, revised version
\vspace{2cm}
\begin{center}
{\bf \huge The Onset of
Color Transparency in $(e,e'p)$ Reactions on Nuclei\\}
\vspace{1cm}
{\bf N.N.Nikolaev$^{1,2)}$, A.Szczurek$^{1,3)}$,
J.Speth$^{1)}$, J.Wambach$^{1,4)}$, \\
B.G.Zakharov$^{5)}$, V.R.Zoller$^{6)}$ } \medskip\\
{\small \sl
$^{1)}$IKP(Theorie), Forschungszentrum  Juelich GmbH.,\\
5170 Juelich, Germany \\
$^{2)}$L.D.Landau Institute for Theoretical Physics, \\
GSP-1, 117940, ul.Kosygina 2, V-334 Moscow, Russia\\
$^{3)}$ Institute of Nuclear Physics, PL-31-342 Krakow, Poland\\
$^{4)}$Department of Physics, University of Illinois at
Urbana-Champaign, \\
Urbana, IL 61801, USA\\
$^{5)}$Nuclear Research Institute, High Energy Laboratory,  \\
Moscow State University, 177234 Moscow, Russia\\
$^{6)}$Institute for Theoretical and Experimental Physics, \\
ul.B.Cheremushkinskaya 25, 117259 Moscow, Russia
\vspace{1cm}\\}
{\bf \LARGE A b s t r a c t \bigskip\\}
\end{center}
Quantum filtering of the ejectile wave packet from hard
$ep$ scattering on bound nucleons puts stringent constraints
on the onset of color transparency in $(e,e'p)$ reactions in
nuclei at moderate energies. Based on multiple-scattering
theory, we derive a novel formula for nuclear transparency
and discuss its energy dependence in terms of a
color transparency sum rule.

%--------------------------------------------------
\newpage
\section{ Introduction.}

There has been  much discussion on
the possibility of color transparency (weak final state interaction,
weak nuclear attenuation) in $(e,e'p)$ reactions on nuclei [1-9].
It is expected that, at very large $Q^{2}$ the nuclear attenuation
of the ejected protons should vanish, as the hard $ep$ scattering
selects a small-size configuration of quarks [1,2]
\beq
\rho \propto 1/Q \,\, ,
\label{eq:1.1}
\endeq
and for the colourless $q\bar{q}$ or  $3q$ systems of
small size $\rho$ the interaction cross section is small [10,11]:
\beq
\sigma(\rho) \propto \rho^{2}  \,\,  .
\label{eq:1.2}
\endeq
Nuclear attenuation will be weak if
the small size (\ref{eq:1.2}) stays frozen while the ejectile
propagates through the nucleus, i.e.,
if the so-called formation (coherence) length [12]
$l_{f}$ is larger than the radius of a nucleus $R_{A}$:
\beq
l_{f}  = \gamma {1\over \Delta m}  \gsim R_{A}  \,\,  ,
\label{eq:1.3}
\endeq
where
$\Delta m \sim (0.5-1.0) GeV$
is a typical level splitting of the nucleonic
resonances and $\gamma$ is the relativistic $\gamma$-factor
of the ejectile state. One then needs for the ejectile proton energy
(which is equal to the energy loss of the scattered  electron) that
$\nu \gsim (3-5)GeV\cdot A^{1/3}$ and that
\beq
 Q^{2} \approx 2m_{N}\nu \gsim (5-10)A^{1/3}(GeV/c)^{2}    \,\,   .
\label{eq:3.1}
\endeq
This shows that in the foreseeable future all the experimental
data will correspond to $l_{f} \sim R_{A}$.

In the practical terms, it is not clear which $Q^{2}$ is large enough
for the onset of the asymptotic shrinkage (\ref{eq:1.1}) [7,8].
Besides, even if the system of quarks emerging from the hard $ep$
scattering vertex has a small size, much depends on how this wave
packet evolves when propagating through the nucleus. Evidently,
the small-size configurations, like any non-stationary state,
should be expanded in terms of the hadronic mass-eigenstates.
Each hadronic state has a {\sl large} free-nucleon cross
section, $\gsim \sigma_{tot}(pN)$, and  weak attenuation can only
come from Gribov's inelastic shadowing [13], i.e., from the
quantum interference of the diagonal (elastic) and off-diagonal
(inelastic) diffractive transitions between the nucleon and
its excited states when the ejectile propagates in the nuclear medium.
All these transitions should be properly phase-correlated. Small
initial size of the ejectile requires many conspiring excited states
and, to keep them all phase-correlated, may require very large
incident energy.

The subject of this paper is the formulation of a quantitative
criterion for the weak final-state interaction in
terms of Gribov's theory of inelastic shadowing [13].
We formulate a "color transparency (CT) sum rule" which
demonstrates clearly the close connection between the onset of
CT and Gribov's inelastic shadowing corrections.
We derive a formula for nuclear transparency with complete
treatment of the quantum interference effects
in the propagation of the ejectile. This has not been done before.
We find that the onset of CT $vs.$ incident energy is very slow and is
very sensitive to the mass spectrum of the diffractively produced
states.

Our analysis is based on the coupled-channel reformulation [14,15]
of Gribov's theory of inelastic shadowing. We follow closely the
treatment of the quasielastic scattering off nuclei as discussed in Ref.~15.

%---------------------------------------------------------------

\section{Color transparency sum rule}

%---------------------------------------------------------------
In order to set a frame of reference, let us consider a two-dimensional
$q\bar{q}$ system with the vector $\vec{\rho}$ being its transverse
size in the plane normal to ejectile's momentum.
The ejectile can be described in terms of two complete sets of wave
functions: the mass eigenstates $|i\rangle$ and the fixed-$\vec{\rho}$
states $|\vec{\rho}\rangle$. The mass-eigenstates $|i\rangle$ can
be expanded in the basis $|\vec{\rho}\rangle$ yielding the stationary
wave function
\beq
\Psi_{i}(\vec{\rho})=\left<\vec{\rho}|i\right> \, .
\label{eq:2.1}
\endeq
Vice versa, the $\vec{\rho}$-eigenstates can be expanded as
\beq
\left.|\vec{\rho}\right>=
\sum_{i}\Psi_{i}(\vec{\rho})^{*}\left.|i\right>     \, .
\label{eq:2.2}
\endeq
Let $\hat{\sigma}$ be the cross section or diffraction operator.
In the $\vec{\rho}$-representation $\hat{\sigma}$ is a diagonal operator:
$\hat{\sigma} = \sigma(\rho)$. In the mass-eigenstate basis the diagonal
matrix elements give the total cross section:
$\sigma_{tot}(iN)=\sigma_{ii}= \left<i|\hat{\sigma}|i\right>$ and
the differential cross section for forward elastic scattering at
$t=0$,
$d\sigma_{el}(iN)/dt\bigr\vert _{t=0}=\sigma_{ii}^{2}/16\pi$.
The off-diagonal matrix elements
$\sigma_{ik}= \left<i|\hat{\sigma}|k\right>$
describe  the differential cross section of the forward diffraction
excitation $k\, N \rightarrow i\, N$:
\beq
\left.{d \sigma_{D}(k N \rightarrow i N) \over dt}\right\vert _{t=0}=
{\sigma_{ik}^{2} \over 16\pi}                            \, .
\label{eq:2.3}
\endeq
The mass-eigenstate expansion for $\sigma(\rho)$ reads  as
\beq
\sigma(\rho)=\left<\vec{\rho}|\hat{\sigma}|\vec{\rho}\right> = \sum_{i,k}
\left<\vec{\rho}|k\right>
\left<k|\hat{\sigma}|i\right>\left<i|\vec{\rho}\right>=
\sum_{i,k} \Psi^{*}_{k}(\vec{\rho})\Psi_{i}(\vec{\rho})
\sigma_{ki}\quad .
\label{eq:2.4}
\endeq
Experimentally, the diffraction excitation rate is much smaller
than the elastic scattering rate,
$\sigma_{ik} \ll \sigma_{ii},\sigma_{kk}$ (see below), so that one
might be tempted to neglect the off-diagonal terms
$\propto \sigma_{ik}$ in (\ref{eq:2.4}). In doing so one runs
into a contradiction with CT as formulated in eq.~(2) since
\beq
\sigma(\rho) \approx \sum_{i}
\Psi^{*}_{i}(\vec{\rho})\Psi_{i}(\vec{\rho}) \sigma_{tot}(i\,N)\sim
\sigma_{tot}(pN)\sum_{i} |\Psi(\vec{\rho})_{i}|^{2}=
\sigma_{tot}(pN)  \,\,  ,
\label{eq:2.5}
\endeq
with the $l.h.s$ expected to vanish as $\rho \rightarrow 0$,
whereas the $r.h.s$ does not depend on $\rho$ at all.

This makes it obvious that, in the mass-eigenstate basis,
CT has its origin in the strong cancellations between the diagonal
and off-diagonal diffractive transitions [8,16]. QCD, as the theory of
strong interactions, dictates a very special relationship between
the wave functions at the origin, $\Psi_{i}(0)$, and the matrix of
the diffraction transition amplitudes $\sigma_{ik}$, which may
be called the "CT sum rule"
\beq
\sum_{i,k} \Psi_{k}(0)^{*}\Psi_{i}(0)\sigma_{ki} = 0      \, .
\label{eq:2.6}
\endeq
The time evolution of the wave packet (\ref{eq:2.2}) is given by
\beq
\left.|\vec{\rho},t\right>=
\sum_{i}\Psi_{i}(\vec{\rho})^{*}\left.|i\right> \exp(-im_{i}t)
\label{eq:2.7}
\endeq
and
\beq
\left<\vec{\rho},t|\hat{\sigma}|\vec{\rho},t\right> = \sum_{i,k}
\left<\vec{\rho}|k\right>\left<k|\hat{\sigma}|i\right>
\left<i|\vec{\rho}\right>
\exp[i(m_{k}-m_{i})t]  \, .
\label{eq:2.8}
\endeq
The proper time $t$ is related to the distance $z$ from the
$ep$ scattering vertex as  $t=z/\gamma$. As soon as large phases
$(m_{k}-m_{i})z/\gamma \sim z/l_{f} \gsim 1$  emerge in
the phase factors of (\ref{eq:2.8}), they will destroy the delicate
cancellations necessary for CT.
Then, the crucial issue is how rapidly the CT sum rule
(\ref{eq:2.6}) is saturated by the lowest-lying excitations
of the nucleon. Saturation requires the excitation of
high-lying nucleonic resonances and therefore impracticably
large $Q^{2}$ may be needed for an onset of CT.

%-------------------------------------------------------------------

\section{Color transparency criterion and diffraction scattering}

%-------------------------------------------------------------------

The CT sum rule (\ref{eq:2.6}) can be given a still more practical
formulation. Since $\sigma(\rho)$ is the eigenvalue of the cross
section operator $\hat{\sigma}$, the CT sum rule simply states that
$\hat{\sigma}$ has a vanishing eigenvalue [17].

The ejectile state $|E\rangle$ emerging from the hard $ep$ scattering
is a coherent superposition of nucleonic states $|i\rangle$
with the relative amplitudes given by the elastic and transition
electromagnetic form factors $G_{ip}(Q^{2})=\langle i| J_{em} |p\rangle$.
Thus
\beq
|E\rangle = \sum G_{ip}(Q^{2})|i\rangle .
\label{eq:4.01}
\endeq
The proton, as the ground state of the mass operator, is expected
to have the smallest size of all the nucleonic states.
In order to form a wave packet $|E\rangle$ of the
size $\rho \ll R_{p}$, where $R_{p}$ is the proton radius,
one has to mix in the higher excitations which have an ever growing
size. According to eq.~(\ref{eq:2.8}), however, the higher
the excitation energy, the faster the corresponding
admixture to $|E\rangle$ becomes out of phase during the evolution,
so that only finite number of excited states $N_{eff}(Q)$ can interfere
coherently at finite energy.

In order to set up a simple formalism for the $Q$(or $\nu$) dependence of
$N_{eff}(Q)$, let us start with the proton-nucleus total cross section.
The conventional Glauber formula reads
\beq
\sigma_{tot}(pA)=2\int d^{2}\vec{b}\,
\left\{ 1 -\exp\left[-{1\over 2}\sigma_{tot}(pN)T(b)\right]\right\}
\label{eq:4.1}
\endeq
and corresponds to the sum of all multiple-scattering diagrams
with the free-nucleon cross section $\sigma_{tot}(pN)$ (Fig.~1a).
Here $b$ stands for the impact parameter,
$T(b)=\int_{-\infty}^{+\infty}dz\,n_{A}(b,z)$
is the familiar optical thickness of the nucleus and $n_{A}(b,z)$
denotes the target matter density.
The inclusion of off-diagonal diffractive transitions shown in
Fig.~1b (Gribov's inelastic shadowing [13]), corresponds to
the coupled-channel generalization of (\ref{eq:4.1}) [14,15]
\beq
\sigma_{tot}(pA)=2\int d^{2}\vec{b}\,
\left\{ 1 -\langle p|
\exp\left[-{1\over 2}\hat{\sigma}T(b)\right]|p\rangle\right\}
\label{eq:4.2}     \,  .
\endeq
In order to see how the high-energy formula (\ref{eq:4.2}) should
be modified at moderate energy, consider the $\nu$-fold scattering
contribution to (\ref{eq:4.2})
\arr
{1\over \nu!}
T(b)^{\nu}\langle p|\hat{\sigma}^{\nu}
|p\rangle =~~~~~~~~~~~~~~~~~~~~~~~~~~~~~~~~~~~~~~~~~~~~~~~~
\nonumber \\
\sum_{i,j...k}\sigma_{pi}\sigma_{ij}...\sigma_{kp}
\int_{-\infty}^{+\infty}dz_{\nu}n_{A}(b,z_{\nu})....
\int_{-\infty}^{z_{2}}dz_{1}n_{A}(b,z_{1})
\label{eq:4.3}
\endarr
Each off-diagonal transition $i\, N \rightarrow k\, N$ involves
a longitudinal momentum transfer
\beq
\kappa_{ik}={ m_{k}^{2} - m_{i}^{2} \over 2E   }
\sim {1 \over l_{f}}\,
\label{eq:4.4}
\endeq
and a phase factor
\beq
\exp\left[i\kappa_{pi}z_{\nu}+ i\kappa_{ij}z_{\nu-1}
+...+i\kappa_{kp}z_{1}\right]
\label{eq:4.5}
\endeq
emerges in the integrand of the $r.h.s.$ of
(\ref{eq:4.3}).
In time-ordered perturbation theory with conserved
momentum $\vec{p}$ one finds the same phase factor when expanding
the energy of the relativistic intermediate state as
$\nu_{i}=\sqrt{p^{2}+m_{i}^{2}} \approx p + m_{i}^{2}/2\nu$ and
factoring out the common  $t$-dependence in
eq.~(\ref{eq:2.7}).

Eq.~(\ref{eq:4.2}) holds at high energies, when $l_{f} \gg R_{A}$, and
the phase in (\ref{eq:4.5}) can be neglected. In this limit a convenient
way to calculate the nuclear cross section is to diagonalize the
diffraction matrix $\hat{\sigma}$ in order to find the corresponding
eigenstates $|\alpha\rangle$ and eigenvalues
$\sigma_{tot}(\alpha\,N)=\langle \alpha|\hat{\sigma}|\alpha\rangle$,
then expand the proton wave function
in these diffraction eigenstates
\beq
|p\rangle = \sum_{\alpha}a_{\alpha}|\alpha\rangle
\label{eq:4.6}
\endeq
and calculate the cross section (\ref{eq:4.2}) as  [14,15]
\beq
\sigma_{tot}(pA)=\sum_{\alpha}a_{\alpha}^{*}a_{\alpha}
2\int d^{2}\vec{b}\,
\left\{ 1 -
\exp\left[-{1\over 2}\sigma_{tot}(\alpha\,N)T(b)\right]\right\}    \,  .
\label{eq:4.7}
\endeq

In contrast, at lower energies, when
$\kappa_{ik}R_{A} \sim R_{A}/l_{f} \gg 1$, contributions
from the Gribov's off-diagonal rescatterings will vanish upon
integration over the nucleon's positions $z_{i}$ and only the
elastic rescattering, $p\,N \rightarrow p\,N$, contribution
survives in the $l.h.s.$ of (\ref{eq:4.5}). As a result,
the nuclear cross section will be given by Glauber's formula
(\ref{eq:4.1}) with the free-nucleon cross section $\sigma_{tot}(pN)$.

At intermediate energies, the result of the $z$ integrations with
the phase factor (\ref{eq:4.5}) in the $r.h.s.$ of eq.~(\ref{eq:4.3})
will be that the $l.h.s$ of eq.~(\ref{eq:4.3}) is multiplied by a certain
form factor $F_{\nu}(b,\kappa_{ij})$. In a somewhat crude approximation,
one can use for this form factor a factorized form
\beq
F_{\nu}(b,\kappa_{ij})\approx \prod^{\nu} G_{A}(\kappa_{ij}) \, ,
\label{eq:4.8}
\endeq
where $G_{A}(\kappa)$ is the charge form factor of the target nucleus
(see, for instance, considerations in [18]).
Then, at each rescattering vertex, the form factor
$G_{A}(\kappa_{ik})$ can be absorbed into the corresponding
rescattering amplitude
\beq
\sigma_{ik}^{eff}=\sigma_{ik}G_{A}(\kappa_{ik})\, .
\label{eq:4.9}
\endeq
Therefore, at moderate energy, one has to solve for the eigenvalues
and eigenstates of the effective energy-dependent diffraction operator
$\hat{\sigma}^{eff}$ and calculate the nuclear cross section
applying eqs.~(\ref{eq:4.6}) and (\ref{eq:4.7}). The factorization
(\ref{eq:4.8}) is an approximation, but it correctly describes
the opening of new coherently interfering channels with
\beq
m_{i}^{2} -m_{p}^{2} \sim {\nu \over R_{A}}
\label{eq:4.10}
\endeq
as the incident energy gradually increases. Notice, that
for elastic rescatterings $\kappa_{ii}=0$ and $G_{A}(\kappa_{ii})=1$.
When the form factor becomes small and the Gaussian approximation
becomes too crude, the corresponding state decouples anyway. Therefore,
using the diffraction operator $\hat{\sigma}^{eff}$ as defined by
eq.~(\ref{eq:4.9}) is justified.

In the $(e,e'p)$ reaction on nuclei the incident particle
is the virtual photon emitted by the electron. In the corresponding
multiple-scattering diagrams of Fig.~1 the rightmost vertex involves
$G_{j}(Q^{2})$. We are interested in those quasielastic
$e+A \rightarrow e+p+A^{*}$ reactions, where one sums over all
excitations of the target debris $A^{*}$, excluding production
of secondary particles (mesons). In the high-energy limit,
the coupled-channel formula for the nuclear transparency or more
precisely the transmission coefficient $Tr_{A}$ (which
can easily be derived following the technique of Ref.~15;
a detailed derivation will be presented elsewhere) takes the form
\beq
Tr_{A} = {d\sigma_{A} \over Z d\sigma_{p} }=
{1 \over |\sum_{\alpha}a_{\alpha}^{*}c_{\alpha}|^{2} }
\sum_{\alpha,\beta}
a_{\alpha}^{*}c_{\alpha}
c_{\beta}^{*}a_{\beta}Tr(\Xi_{\alpha\beta})       \,  ,
\label{eq:5.3}
\endeq
where
\beq
Tr(\sigma)={1 \over A\sigma}\int d^{2}\vec{b}
\left\{1-\exp\left[-\sigma T(b)\right]\right\}    \, ,
\label{eq:5.4}
\endeq
\beq
\Xi_{\alpha\beta}=
{1 \over 2}(\sigma_{tot}(\alpha\,N) + \sigma_{tot}(\beta\,N)-
\int d\Omega |f_{\alpha}(\theta)f_{\beta}(\theta)|
\label{eq:5.5}
\endeq
and $c_{\alpha}$ is defined by expansion of the ejectile in the
eigenvalues of the diffraction operator $\hat{\sigma}$
\beq
| E\rangle = \sum_{\alpha} c_{\alpha}|\alpha\rangle .
\label{eq:5.6}
\endeq
In (\ref{eq:5.5}) $f_{\alpha}(\theta)$ is the corresponding elastic
scattering amplitude ($\sigma_{el}=$$\int d\Omega |f_{el}(\theta)|^{2}$).
Notice, that the numerator of eq.~(\ref{eq:5.4}) is precisely the
inelastic nuclear cross section $\sigma_{in}(hA)$ for the free-nucleon
cross section $\sigma_{tot}(hN)=\sigma$.
The emergence of the non-trivial quantity $\Xi_{\alpha\beta}$
demonstrates the importance of the quantal interference effects in the
multiple-scattering problem [15].

At moderate energies, one has to calculate the nuclear
transparency (\ref{eq:5.3}) using the diffraction operator
$\hat{\sigma}^{eff}$. Simultaneously, one has to apply
the rule (\ref{eq:4.9}) to the electromagnetic vertex as well:
$G_{kp}(Q^{2}) \rightarrow G_{kp}(Q^{2})G_{A}(\kappa_{kp})$.

In the low-energy limit of $l_{f} \ll R_{A}$ all inelastic
excitations decouple and the coupled-channel problem reduces
to a single-channel problem.
In this case $\sigma_{tot}(\alpha\,N)=\sigma_{tot}(pN)$, the last term
in (\ref{eq:5.5}) is the elastic cross section $\sigma_{el}(pN)$
and $Tr_{A}$ is given by eq.~(\ref{eq:5.4}) with
$\sigma = \sigma_{in}(pN)$. This result is quite obvious, since
the elastic rescatterings of the high-energy ejectile do not
contribute to nuclear attenuation. Glauber and Matthiae have
noticed this in 1970 in their multiple-scattering analysis [19]
of the quasielastic proton-nucleus scattering (see also
a recent paper by Kohama, Yazaki and Seki [9]).

%----------------------------------------------------------------

\section{Estimates for nuclear transparency}

%----------------------------------------------------------------

A possible signal of CT in $(e,e'p)$ reactions depends on how
baryons are excited in both
the hard electromagnetic and the soft diffractive scattering, i.e.,
on the internal structure of the nucleon. In high-energy QCD,
the quark-spin
changing diffractive transitions are weak. Hence the candidate states
which could conspire to produce an ejectile of small {\sl transverse}
size ejectile are the $N(938,{1\over 2}^{+})$,$N^{*}(1680,{5\over 2}^{+})$,
$N^{*}(2220,{9\over 2}^{+})$ as well as higher excitations.
(In the two-dimensional case only the radial excitations contribute
to the CT sum rule, in the three-dimensional case angular excitations
to the same parity states contribute as well).
Excitation of the $N^{*}(1680,{5\over 2}^{+})$ is one of the
prominent channels of the diffraction-dissociation of nucleons with
$\sigma(NN\rightarrow
N^{*}(1680,{5\over 2}^{+})) \approx 0.17 mb$ [20,21] as
compared to $\sigma_{el}(NN\rightarrow NN) \approx 7mb$.
The total inclusive cross section of diffractive excitations sums up
to $ \sim 2mb$ [21].

As we have discussed above, the requirement of phase coherence,
limits the number of interfering states $N_{eff}(Q)$
from above. The case of $N_{eff}=1$ corresponds to
$\sigma_{min} = \sigma_{11}=\sigma_{tot}(pN)$ and represents
the conventional nuclear attenuation. The case of $M=2$ gives
\beq
\sigma_{min}= {1 \over 2}(\sigma_{11}+\sigma_{22}) -
{1 \over 2}\sqrt{ (\sigma_{11}-\sigma_{22})^{2}+4\sigma_{12}^{2}}
\label{eq:3.3}
\endeq
so that $\sigma _{min} \geq \sigma_{tot}(NN)-|\sigma_{12}|$.
Judging from the magnitude of the diffraction-dissociation cross
sections, one cannot gain much in transmission when the number of
interfering states is small. Two states are simply not enough to
form a wave packet of small size and small $\sigma_{min}$
(excitation and coherent interference of many states is indeed
important for CT).

There exists considerable evidence that the nucleon
can be regarded as a quark core surrounded by a pionic
cloud [22] which, in the simplest approximation, yields a
wave function
\beq
 |N\rangle = \cos(\theta)|3q\rangle + \sin(\theta)|3q +\pi\rangle \,  .
\label{eq:6.01}
\endeq
{}From the diffractive scattering point of view, one can estimate the
core size from
$\sigma_{tot}(NN) \approx \sigma_{tot}((3q)N) +
n_{\pi}\sigma_{tot}(\pi N) $, where $n_{\pi}$ is the average number
of pions in the nucleon.
Here we have neglected the shadowing in the $3q+\pi$ system. With
$\sigma_{tot}(\pi N) \approx 2/3 \sigma_{tot}(NN)$ we then have
\beq
\sigma_{tot}((3q)N) \approx
(1-{2 \over 3}n_{\pi})\sigma_{tot}(NN)
\label{eq:6.02}
\endeq
which gives a lower bound for $\sigma_{tot}((3q)N)$. Evaluation
of the number of pions in the nucleon with a realistic $\pi N$ vertex
function gives $n_{\pi} \sim 0.4$ [22].
The same admixture of pions describes diffraction production of
the continuum $N\pi$ system by the so-called Drell-Hiida-Deck
mechanism (for a review see [21]).

{}From the point of view of electromagnetic scattering, the $3q+\pi$
admixture has a large radius and contributes significantly to
the charge radius of the proton. However, because of the large size,
the contribution of the  $3q+\pi$ component to the nucleon
electromagnetic form factor vanishes rapidly at large $Q^{2}$.
Our estimates with realistic pion-nucleon vertex functions show
that the 3-quark Fock state dominates already at
$Q^{2} \gsim 1 (GeV/c)^{2}$. The electromagnetic form factor
analysis suggests $\langle R_{ch}^{2} \rangle_{3q} \approx
0.8 \langle R_{ch}^{2} \rangle_{p}$, which is consistent with
eq.(\ref{eq:6.02}).

Therefore, at $Q^{2} \sim 10 (GeV/c)^{2}$, the region of  interest
for CT, we can concentrate on excitations of the 3-quark core. In order
to study the onset of color transparency, we calculate the diffraction
matrix elements $\sigma_{ik}$ in the quark-diquark model of the nucleon,
using three-dimensional harmonic oscillator wave functions and choose
the $\rho$ dependence of the interaction cross section as
\beq
\sigma(\rho)=\sigma_{0}
\left[1-\exp\left(-{\rho^{2}\over R_{0}^{2}}\right)\right] .
\label{eq:6.1}
\endeq
This form reproduces the color transparency property (\ref{eq:1.2})
and saturates at large $\rho$ in agreement with confinement [23].
Following the calculations in Ref.~[23], we take
$\sigma_{0} \approx 1.6\sigma_{tot}(pN)$ and
$R_{0}^{2} \approx 0.55\, fm^{2}$, such that the $3q$-core interacts
with the nucleon with the cross section $\approx 0.8 \sigma_{tot}(pN)$,
cf. eq.~(\ref{eq:6.02}).
In a model of scalar diquarks and with the QCD suppression of
the spin-flip transitions, the dominant diffraction transitions are
to the even angular momentum and positive parity quark-diquark states
with the angular momentum projection $L_{z}=0$. Production of
$N^{*}(1680,{5\over 2}^{+})$ is a prominent feature of the forward
diffraction dissociation of nucleons, which suggests
$\Delta m =2\hbar \omega \approx 700 MeV/c^{2}$. The radial excitations
of nucleons in the non-relativistic quark model rather suggest
$\Delta m \approx 1200 MeV/c^{2}$. This may be a more realistic
estimate if we ask for a large number of interfering states
$N_{eff}$ ($N_{eff}=1,3,6,10$ with the number of interfering major shells
$K=1,2,3,4$). For our purpose it is sufficient to take a Gaussian
nuclear form factor
$G_{A}(\kappa)=\exp(-{1 \over 6}\langle R_{ch}\rangle_{A}^{2}\kappa^{2})$
when evaluating the effective diffraction matrix $\hat{\sigma}^{eff}$.

In Table 1 we show the structure of the diffraction matrix
$\hat{\sigma}$. Its off-diagonal matrix elements give realistic
estimates for the diffraction dissociation cross sections
$\sigma_{D}(pN\rightarrow ip) \sim
(\sigma_{tot}(iN)/\sigma_{tot}(pN))^{2} \sigma_{el}(pp)$.
In Table 2 we show the eigenvalues of the truncated diffraction matrix
$\hat{\sigma}$. The departure of the minimal eigenvalue $\sigma_{min}$
from zero measures the rate of saturation of the CT sum rule in
the limited basis of $N_{eff}$ states. The minimal eigenvalue
$\sigma_{min}(Q)$ decreases at large $N_{eff}(Q)$, but this decrease
is very slow (see below). Fig.2 displays the lowest eigenvalue
of $\hat{\sigma}^{eff}$ as a function of energy for few nuclei.
The corresponding effective number of interfering shells $K$ can
easily be deduced from Table 2. The $Q$($\nu$) dependence of
$\sigma_{min}$ can be understood from the condition
(\ref{eq:4.10}). For instance, at $\nu \sim 30\, GeV$ only the ground
state (proton, $K=1$) and the two excited shells, $K=2,3$,
can interfere coherently in the $(e,e'p)$ reaction on the carbon nucleus.
Notice, that excitation of the same number of shells on  different nuclei
requires an energy $\nu \propto A^{1/3}$, and we predict therefore slower
onset of reduced nuclear attenuation for heavy nuclei as compared
to higher nuclei.

We can optimize for the transparency by requiring that the ejectile
state $|E \rangle$ coincides with the eigenfunction corresponding
to the smallest eigenvalue $\sigma_{min}$ of the diffraction matrix.
The resulting transparency will be given by formula (\ref{eq:5.4}) with
\beq
\sigma \approx \sigma_{min}\left(1-
{\sigma_{min} \over \sigma_{tot}(pN) }
{\sigma_{el}(pN)\over \sigma_{tot}(pN) }\right)     \,  .
\label{eq:7.3}
\endeq
For instance, when $K=1,2,3$ shells are included, the eigenstate
with $\sigma_{min}=17.5 mb$, appropriate for an energy $\nu \sim 30 GeV$,
corresponds to the superposition of the harmonic oscillator states
$|n_{r},L\rangle$
\arr
|1,0\rangle\,:\,
|2,0\rangle\,:\,
|1,2\rangle\,:\,
|3,0\rangle\,:\,
|2,2\rangle\,:\,
|1,4\rangle  =   \nonumber \\
0.77\,:\,0.44\,:\,0.31\,:\,0.26\,:\,0.22\,:\,0.11~~~~~~ \,  .
\label{eq:7.4}
\endarr
For the radial excitations, the asymptotics of the non-relativistic
form factor will be roughly proportional to the product of wave
functions at the origin, and the transition form factors will be of
the same order as the elastic scattering form factor, in agreement with
the expansion given by (\ref{eq:7.4}) and with the conclusions by Stoler
from an analysis [24] of the SLAC electroproduction data.
With two shells ($\sigma_{min}=22.1 mb$) and with four shells
($\sigma_{min}=14.0 mb$) the corresponding eigenstates are
\beq
|1,0\rangle\,:\,
|2,0\rangle\,:\,
|1,2\rangle
 =
0.85\,:\,0.43\,:\,0.30       \,  ,
\label{eq:7.5}
\endeq
\arr
|1,0\rangle\,:\,
|2,0\rangle\,:\,
|1,2\rangle\,:\,
|3,0\rangle\,:\,
|2,2\rangle\,:\,
|1,4\rangle\,:\,
|4,0\rangle\,:\,
|3,2\rangle\,:\,
|2,4\rangle\,:\,
|1,6\rangle  =   \nonumber \\
0.70\,:\,0.43\,:\,0.31\,:\,0.28\,:\,0.24\,:\,0.18\,:\,0.17\,:\,
0.10\,:\,0.04~~~~~~~~~~~~
\label{eq:7.6}
\endarr

In Fig.3 we show the energy dependence of the nuclear transparency
factor $Tr_{A}$ for these least-attenuation states. As a reference
point one should take the low-energy values of transparency evaluated
from formula (\ref{eq:5.4}) with
 $\sigma=\sigma_{in}(pN)\approx 30 mb$: ~
$Tr_{C}=0.61$, ~~$Tr_{Al}=0.49$,~~$Tr_{Cu}=0.40$,~~$Tr_{Pb}=0.26$,
which are also shown in Fig.3. We remind that the
curves shown correspond to the $3q$-core stripped off the pionic
cloud and having $\sigma_{tot}((3q)N) \approx 0.8\sigma_{tot}(pN)$.
At $\nu \sim 15 GeV$ we find $ \sim 35\%$ reduction of attenuation for
carbon, and $ \sim 25\%$ effect for aluminum. The effect decreases
significantly if one chooses $\Delta m \approx 1.2 GeV/c^{2}$.

Notice, that with the optimized ejectile state $|E\rangle$, the
nuclear transparency in $(e,e'N^{*})$ reactions is the same as in
$(e,e'p)$ reaction, although for resonances the free-nucleon cross
section is larger than for protons, cf. the diagonal matrix elements
in Table 1. This equality holds at high enough energies, when
for excitation of the corresponding shell $\kappa_{N^{*}p}R_{A} \ll 1$.

As we have explained above, states of a large number of the oscillator
shells must conspire to produce a wave packet of very small size.
However, at energies of practical interest $\nu \sim (10-30) GeV$ and
$Q^{2} \sim (20-60) (GeV/c^{2})$ the admixture of shells with
$K \gsim 4$, even if present in the initial ejectile state $|E\rangle$ ,
will get very rapidly out of phase and cannot enhance the nuclear
transparency above our estimates shown in Fig.3.

The principal parameter which controls the onset of CT is the size
of the $3q$-core, which is sensitive to the number of pions in
nucleons, see also eq.(\ref{eq:6.02}). If
$\langle R_{ch}^{2} \rangle_{3q} = 0.5 \langle R_{ch}^{2} \rangle_{p}$,
then $\sigma_{tot}((3q)N) = 24 mb$ and, compared to the previous
case, $\sigma_{min}$ will be $\sim 25\%$ smaller. We then find
a somewhat stronger signal of transparency at large $Q^{2}$.

The above conclusions hold for the electroproduction of pions as well,
where the first excited shell corresponds to the
$\pi(1300,0^{-})$,$A_{1}(1260,1^{+})$ and $\pi_{2}(1670,2^{-})$ resonances.
In this case the appropriate energy scale will be set by
$\Delta m \sim 1.2 GeV/c^{2}$.

%----------------------------------------------------------------

\section{Discussion of the results}

%----------------------------------------------------------------

We have constructed a realistic model of CT effects in $(e,e'p)$
reactions on nuclei. Our formalism, embodied in the target-size
and energy-dependent diffraction matrix (\ref{eq:4.8})
and in the nuclear transparency formulae (\ref{eq:5.3})-(\ref{eq:5.5}),
correctly describes the quantum interference effects in the ejectile
propagation through the nucleus. Even after optimizing the ejectile
state scattering for weak attenuation,
we find rather slow onset of CT. This can be understood in terms of
a large number of excited states needed to produce wave packets of
small transverse size. At moderate energies the coherent interference
condition constrains $N_{eff}(Q)$ from above. Nevertheless, for
light nuclei (carbon or aluminum) the predicted CT signal is
strong enough for a decisive test of CT ideas at the highest
energy at SLAC or at a new European Electron Facility currently
under discussion.
CT experiments in $(e,e'p)$ scattering are much discussed as a case for
electronuclear facilities of next generation [1-9,25-28].
Our analysis strongly suggests that electron beams of $\sim 30 \,GeV$
are highly desirable in order to observe a CT signal for heavy
nuclear targets.

Major differences from early papers on CT effects in $(e,e'p)$
scattering is as follows. Farrar, Frankfurt, Liu and Strikman [3]
(see also [4]) use a classical expansion and attenuation model,
which is too crude to describe the quantal interference effects.
This has been discussed in Ref.~[29] for photoproduction of charmonium
off nuclei. The approach of Jennings and Miller [5] is
similar to ours to the extent that they also start from the
multiple scattering theory. Their QCD-motivated model
for the diffraction operator, used in [5], satisfies our CT sum rule.
We differ, however, significantly in the treatment of the coherence
constraint.  While we treat it within the coupled-channel
theory, Jennings and Miller reduce the coupled-channel
problem to a single-channel problem with a complex cross section.
After this paper was submitted for publication, we have learned of
preliminary results of the NE18 experiment at SLAC, which confirm
our predictions of a  slow onset of CT in $(e,e'p)$ scattering [30].
\vspace{2.0cm}

\noindent
{\bf Acknowledgements:}
We are grateful to G.E.Brown and K.Yazaki for discussions.
Two of the authors (N.N.N. and A.S.) thank KFA-IKP(TH) for
the hospitality and financial support. This work was supported in part
by NSF grant PHY-89-21025 and by the Polish KBN grant 2 2409 91 02.
\newpage

% = = = = = = = = = = = = = = = = = = = = = = = = = = = = = = = = = = =

% = = = = = = = = = = = = = = = = = = = = = = = = = = = = = = = = = = =

\newpage

\begin{table}[h]
\center\begin{tabular}{|c||c|c|c|c|c|c||} \hline
$(n_{r},L)$ & $(1,0)$ & $(2,0)$ &  $(1,2)$ & $(3,0)$
& $(2,2)$ & $(1,4)$\\
\hline \hline
$(1,0)$   &  32.0  & -13.1  &  -9.2   &  -5.8  &  -4.9  &  -2.3 \\
\hline
$(2,0)$   & -13.1  &  42.7  &   7.5   & -11.9  &  -4.0  &   4.8  \\
\hline
$(1,2)$   &  -9.2  &  7.5   &   37.3  &  1.7   &  -7.1  &   -12.9 \\
\hline
$(3,0)$   &  -5.8  &  -11.9 &   1.7   &  46.9  &  7.2   &   -2.6  \\
\hline
$(2,2)$   &  -4.9  &  -4.0  &   -7.1  &  7.2   &  40.4  &  6.5 \\
\hline
$(1,4)$   &  -2.3  &  4.8   &  -12.9  & -2.6   &  6.5   &  44.7 \\
\hline \hline
\end{tabular}
\caption{\sl The diffraction matrix (matrix elements are
in millibarns).}
\label{matrix}
\end{table}

\begin{table}
\center\begin{tabular}{|c|c|c|c|} \hline
Number of & Number of the& Excitation & $\sigma_{min}$ \\
the shells, $K$ & coupled states, $M$ & energy (GeV) & (mb) \\
\hline
 1 & 1 & 0.94 & 32.0\\
 2 & 3 & 1.64 & 22.1 \\
 3 & 6 & 2.34 & 17.1 \\
 4 & 10 & 3.04 & 14.0 \\
 5 & 15 & 3.74 & 11.8 \\
 6 & 21 & 4.34 & 10.3\\
 7 & 28 & 5.04 &  9.1\\
\hline
\end{tabular}
\caption{\sl The minimal eigenvalue of the diffraction matrix
vs. the number of the coupled shells $K$
($\Delta m = 0.7 GeV/c^{2}$).}
\label{sigmin}
\end{table}

\newpage

\noindent
{\bf Figure captions}
\bigskip\\

\noindent
Fig.1 - Elastic ($a$) and inelastic ($b$) shadowing in $pA$ scattering.
\bigskip\\
\noindent
Fig.2 - The minimal eigenvalue $\sigma_{min}(\nu)$ $(\nu=Q^2/2m_N)$
of the diffraction
matrix (\ref{eq:4.9}) for the $3q$-core of the nucleon
as a function of $\nu$ for different nuclei
($\langle R_{ch}^{2}\rangle_{3q}=0.8\langle R_{ch}^{2}\rangle_{p}$
and $\Delta m = 0.7 GeV/c^{2}$).
\bigskip\\
\noindent
Fig.3 - The ejectile energy ($\nu = Q^{2}/2m_{N}$) dependence of
the nuclear transparency $Tr_{A}$ in the quasielastic $(e,e'p)$
scattering on different nuclei at $\Delta m = 0.7 GeV/c^{2}$
(solid curves) and $\Delta m = 1.2 GeV/c^{2}$
(dotted curve for the carbon nucleus) for the $3q$-core with
$\langle R_{ch}^{2}\rangle_{3q}=0.8\langle R_{ch}^{2}\rangle_{p}$.
The dashed curves for the carbon and lead nuclei are for
$\langle R_{ch}^{2}\rangle_{3q}=0.5\langle R_{ch}^{2}\rangle_{p}$
and $\Delta m = 0.7 GeV/c^{2}$.
The low energy predictions are shown separately.

\end{document}